\documentclass[twocolumn,preprintnumbers,prb]{revtex4}

\usepackage{amssymb}
\usepackage{amsmath}
\usepackage{graphicx}

\newcommand{\dx}{\mbox{$\Delta r$}}

\begin{document}

\title{Solution of the Percus-Yevick equation for hard hyperspheres in even dimensions}
\author{M. Adda-Bedia\dag, E. Katzav\dag, and D. Vella\dag\ddag}
\affiliation{\dag Laboratoire de Physique Statistique de l'Ecole Normale Sup\'erieure, CNRS UMR8550,
24 rue Lhomond, 75231 Paris Cedex 05, France.\\
\ddag ITG, Department of Applied Mathematics and Theoretical Physics, University of Cambridge, Wilberforce Road, Cambridge CB3 0WA, United Kingdom.}
\date{\today}

\begin{abstract}

We solve the Percus-Yevick equation in even dimensions by reducing
it to a set of simple integro-differential equations. This work
generalizes an approach we developed previously for hard discs. We
numerically obtain both the pair correlation function and the virial
coefficients for a fluid of hyper-spheres in dimensions $d=4,6$ and
$8$, and find good agreement with available exact results and
Monte-Carlo simulations. This paper confirms the alternating
character of the virial series for $d \ge 6$, and provides the first
evidence for an alternating character for $d=4$. Moreover, we show that this sign alternation is due to the existence of a branch point on the negative real axis.
It is this branch point that determines the radius of convergence of
the virial series, whose value we determine explicitly for
$d=4,6,8$. Our results complement, and are consistent with, a recent
study in odd dimensions [R.D. Rohrmann et al., J. Chem. Phys. 129,
014510 (2008)].

\end{abstract}

\maketitle

\section{Introduction}

In recent years, there has been much interest in the properties of
hard sphere fluids. The study of such systems has played a central
role in the understanding of classical fluids and serves as a
starting point for the construction of perturbation theories of
fluid properties. A particular interest has been systems of hard
hyperspheres
\cite{whitlock,Skoge06,Lue05,Lue06,clisby06,B4,B3,clisby04b,clisby05,Rohrmann,rohrmann08,robles,frisch}
(i.e. the generalization of spheres to dimensions larger than three,
$d>3$). There are several reasons for this interest. Firstly, as in
a variety of interacting many-body systems \cite{CL}, one expects
studies of hard-sphere packings in high dimensions to yield great
insight into the corresponding phenomena in lower dimensions, such
as ground and glassy states of matter \cite{Torquato02,Torquato06}.
Secondly, it is hoped\cite{frisch} that analytical investigations of
hard hyperspheres in large spatial dimensions can serve as an
organizing device for a systematic expansion in inverse powers of
the dimension, $d$. From a completely different perspective, hard
hypersphere systems play an important role in communication theory.
For example, it is known that the optimal way of sending digital
signals over noisy channels correspond to the densest sphere packing
in a high-dimensional space \cite{CS}, so called ``spherical codes".

It is common to describe hard-core fluids using approximate
theories\cite{hansen}. There are two main reasons for using such
approximate theories: on the one hand, a solution of the full
problem was, and still is, extremely difficult. On the other hand,
approximate methods have led to very good predictions in the
low density phase. Among the most widely used approximations is the
Percus-Yevick (PY) equation for $d$-dimensional hard spheres
\cite{PY}, which is exact to first order in the density of the
fluid\cite{rohrmann08}, $\rho$.

A great deal of progress has been made towards understanding the
solutions of the PY equation in odd dimensions. In principle, the
work of Baxter\cite{baxter} and Leutheusser\cite{leuth1} reduces the
PY equation to a set of nonlinear algebraic equations of order
$2^{(d-3)/2}$ for $d>3$. However, these can only be solved
analytically for $d \le 7$ so that results in higher odd dimensions
must be found numerically. A more complete survey of the literature
for odd dimensions may be found elsewhere\cite{rohrmann08}.

In even dimensions the situation is much less developed. In two
dimensions an approximate numerical solution of the PY equation was
found by Lado~\cite{lado}. Leutheusser \cite{leuth2} was able to fit
many of Lado's results using an ansatz for the direct correlation
function. Other results available in the literature for hard discs
are solutions of the full problem based on Molecular Dynamics (MD)
or Monte Carlo (MC) methods
\cite{hoover67,numerics,alder68,wood,clisby06}. Recently,
the authors solved the PY equation for hard discs \cite{PY2d}, by
developing a method that reduces the problem to a set of integral
equations that are solved numerically without major difficulties.

In larger even dimensions there are some MD
simulations\cite{Lue05,Lue06,Skoge06}, MC
simulations\cite{whitlock}, MC calculations
\cite{clisby04b,clisby05,clisby06} and a few analytical results for the
low order virial coefficients \cite{B3,B4}. On the analytical front,
Rosenfeld \cite{rosenfeld} generalized Leutheusser's
Ansatz\cite{leuth2} to higher dimensions and compared the results
with the analytical results in three and five dimensions. However,
to our knowledge, the PY equation has not been solved in any even
dimension apart from $d=2$.

In this paper, we solve the PY equation for some even dimensions
($d=4,6$ and $8$). We do so by generalizing our previous work
\cite{PY2d}, which was based on techniques borrowed from the
resolution of crack problems \cite{sned} and uses some results from
Baxter's classical method~\cite{hansen,baxter}. The main difference
from the hard sphere case (and any odd dimension in general) is that
the problem of finding the total correlation function and the direct
correlation function are coupled. This means that the present
analysis necessarily yields both correlation functions and therefore
provides the equation of state. The advantage of the current method
over previous approaches is that it provides all the quantities of
interest as power-series in the density. Thus, questions like
existence of negative virial coefficients in four
dimensions\cite{clisby04b} and more generally the radius of
convergence of the series in large dimensions \cite{rohrmann08} can
be tackled.

\section{The Percus-Yevick approximation}

The pair correlation function $g({\mathbf r})$ is related to the
direct correlation function $c({\mathbf r})$ through the
Ornstein-Zernike equation by \cite{hansen,OZ}

\begin{equation}
h({\mathbf r})=c({\mathbf r})+\rho \int_{0}^{\infty} h({\mathbf
r'})c(|\mathbf{ r}-\mathbf{ r'}|) dr'
 \label{eq:O&Zr} \, ,
\end{equation}
where $\rho$ is the particle number density and
\begin{equation}
h({\mathbf r})=g({\mathbf r})-1\,,
\label{hdefn}
\end{equation}
is the total correlation function\cite{hansen}. The PY approximation
is a closure relation for Eq.~(\ref{eq:O&Zr}). For a hard-core pair
interaction potential, this approximation reads \cite{hansen}
\begin{eqnarray}
g(r) & = & h(r)+1=0, \qquad r<1
\label{eq:bc1}\\
c(r) & = & 0,\qquad r>1
\label{eq:bc2}
\end{eqnarray}
Interestingly, the PY approximation can be seen as a Random Phase
Approximation (RPA) to some nonlinear field theory as shown previously
\cite{SE}.

Here and elsewhere, we take the diameter of the hypersphere to be
unity. Thus in $d$ dimensions ($d=2(k+1)$, with $k\geq0$), we have
\begin{equation}
 \rho= \frac{ \eta}{V_d(1/2)}=\left(\frac{4}{\pi}\right)^{k+1} (k+1)!\, \eta\,
 \label{eq:rho-eta} \, ,
\end{equation}
where
\begin{equation}
V_d(R) = \frac{\pi^{\frac{d}{2}}
R^d}{\Gamma\left(\frac{d+2}{2}\right)}, \label{vddefn}
\end{equation} is the volume of a $d$-dimensional hypersphere of radius $R$, $\eta$
is the packing fraction, and the space filling density corresponds
to $\eta=1$. We define the $d$-dimensional Fourier Transform
\begin{equation}
\label{eq:fourier1}
\tilde{f}(q)=(2\pi)^{k+1}q^{-k}\int_{0}^{\infty}r^{k+1}J_{k}(qr)f(r)dr\,,
\end{equation}
where $J_k$ is the Bessel function and the inverse Fourier transform for $f(r)$ and $F_k(r)$ are given by
\begin{equation}
f(r)=(2\pi)^{-(k+1)}r^{-k}\int_{0}^{\infty}q^{k+1}J_{k}(qr)\tilde{f}(q)dq\,,
\label{eq:fourier1k}
\end{equation}

Applying the $d$-dimensional Fourier transform to
Eq.~(\ref{eq:O&Zr}) yields
\begin{equation}
\tilde{h}(q)=\tilde{c}(q)+\rho \tilde{h}(q)\tilde{c}(q)
 \label{eq:O&Z} \, ,
\end{equation}
which can be rewritten as
\begin{equation}
\left[1-\rho \tilde{c}(q)\right]\left[1+\rho \tilde{h}(q)\right]=1
\label{eq:O&Z1} \,.
\end{equation}

Finally, the static structure factor $s(q)$ of wavenumber ${\mathbf q}$
is related to the pair correlation function through
\begin{equation}
s(q)=1+\rho \tilde{h}(q)\equiv\frac{1}{1-\rho \tilde{c}(q)}
 \label{eq:sq}\,.
\end{equation}

\section{Resolution}

The condition (\ref {eq:bc2}) together with (\ref {eq:fourier1k})
imposes that $\tilde{c}(q)$ can be written without loss of
generality as
\begin{equation}
\tilde{c}(q)=\left(\frac{2\pi}{q}\right)^{k+\frac{1}{2}}\int_0^1 t^{k+\frac{3}{2}}J_{k+\frac{1}{2}}(qt)\phi(t) dt
\label{eq:solcq}
\end{equation}
where $\phi(t)$ is a real function. Substituting \eqref{eq:solcq} into \eqref {eq:fourier1k} and simplifying by using the integral~{\bf 6.575-1} in \cite{grads}, we find that $c(r)$ can
be expressed as
\begin{equation}
 c(r)= \int_r^1 \frac{t\phi(t)}{\sqrt{t^2-r^2}} \frac{dt}{\pi} \, , \qquad 0\leq r<1
 \label{eq:solc} \, .
\end{equation}
Since $c(r)$ is discontinuous yet finite at $r=1$, one has
\begin{equation}
\phi(t)\sim \left(1-t^2\right)^{-1/2}\qquad \mbox{as}\qquad t\rightarrow 1^-
\end{equation}

In the following, we use the
formulation of Baxter for the odd dimensional case, with the
important difference that instead of solving for $c(r)$ directly, we
solve for $\phi(t)$, and obtain $c(r)$ using Eq.~(\ref{eq:solc}). We
use the Wiener-Hopf method by defining
\begin{equation}
 A(q) \equiv \frac{1}{s(q)} = 1-\rho\left(\frac{2\pi}{q}\right)^{k+\frac{1}{2}}\int_0^1
 t^{k+\frac{3}{2}}J_{k+\frac{1}{2}}(qt)\phi(t) dt
 \label{eq:A}
\end{equation}
One sees that $A(q)=A(-q)$, $A(q)\rightarrow 1$ as $q\rightarrow
\infty$ and that $A(q)$ has the same properties as the corresponding
function defined in the odd dimensional case: it has neither zeros
nor poles on the real axis, since by definition $s(q)$ has neither
zeros nor poles for all $q$'s. Therefore one can use the Wiener-Hopf
decomposition of Baxter \cite{hansen,baxter}
\begin{equation}
A(q)=\tilde{Q}(q)\tilde{Q}(-q)
\label{adefn}
\end{equation}
where $\tilde{Q}(q)$ is analytic for $\Im(q)>0$. Following the same steps as
in \cite{hansen,baxter,PY2d} one can show that $\tilde{Q}(q)$ can be
written as
\begin{equation}
 \tilde{Q}(q)=1-\lambda \int_0^1Q(t)e^{iqt}dt
 \label{eq:Qq}
\end{equation}
where $\lambda$ is a parameter defined by
\begin{equation}
\lambda\equiv(2\pi)^k \rho
\label{lambdadefn}
\end{equation}
and thus using \eqref{eq:A}-\eqref{eq:Qq}
\begin{eqnarray}
&\rho& \left(\frac{2\pi}{q}\right)^{k+\frac{1}{2}}\int_0^1 s^{k+\frac{3}{2}}J_{k+\frac{1}{2}}(qs)\phi(s) ds \nonumber \\
&=& \lambda \int_0^1Q(s)e^{iqs}ds+\lambda \int_0^1Q(s)e^{-iqs}ds \nonumber \\
&-& \lambda^2 \int_0^1ds\int_0^1ds'Q(s)Q(s')e^{iq(s-s')}
\end{eqnarray}
Multiplying by $\exp(-iqt)$, with $0 \leq t\leq 1$ and integrating
with respect to $q$ from $-\infty$ to $\infty$ gives
\begin{eqnarray}
&& \sqrt{\frac{2}{\pi}} \int_0^1  ds \, s^{k+\frac{3}{2}}\phi(s) \int_0^\infty dq\, q^{-(k+\frac{1}{2})} J_{k+\frac{1}{2}}(qs) \cos qt \nonumber \\
&& = Q(t)-\lambda \int_t^1 Q(s)Q(s-t)ds
\end{eqnarray}
which can be simplified to give
\begin{equation}
\int_t^1 (s^2-t^2)^k \phi(s)sds= 2^{k}k!\left[Q(t)-\lambda\int_t^1 Q(s)Q(s-t)ds\right].
\label{eq:solphi0}
\end{equation}
Differentiating $k$-times with respect to $t^2$ and once with respect to $t$ gives
\begin{equation}
 \phi(t)= (-1)^{k+1}\left(\frac{d}{tdt}\right)^{k+1}\left[Q(t)-\lambda\int_t^1 Q(s)Q(s-t)ds\right]
 \label{eq:solphi} \, ,
\end{equation}
which is valid for $0\leq t\leq 1$. Therefore, once $Q(t)$ is known, $\phi(t)$ is
given by Eq.~(\ref{eq:solphi}) and $c(r)$ is given by
Eq.~(\ref{eq:solc}).

Now let us work on the function $h(r)$. Since $\tilde{h}(q)$ is an
even function one can write without loss of generality
\begin{equation}
 \tilde{h}(q)=\left(\frac{2\pi}{q}\right)^{k+\frac{1}{2}}\int_0^\infty t^{k+\frac{3}{2}}J_{k+\frac{1}{2}}(qt)\psi(t) dt
 \label{eq:h}
\end{equation}
where $\psi(t)$ is a real function. Then, using
Eq.~(\ref{eq:fourier1k}) $h(r)$ can be written in terms of $\psi(t)$
as
\begin{equation}
 h(r)=\int_r^\infty \frac{t\psi(t)}{\sqrt{t^2-r^2}} \frac{dt}{\pi} \, , \qquad r>0
 \label{eq:solh} \, .
\end{equation}
Combining this equation together with the condition
(\ref{eq:bc1}), one obtains
\begin{equation}
 \int_r^1 \frac{t\psi(t)}{\sqrt{t^2-r^2}} \frac{dt}{\pi} =-1-\int_1^\infty \frac{t\psi(t)}{\sqrt{t^2-r^2}} \frac{dt}{\pi} \, ,  \qquad 0<r<1
 \label{eq:cond1} \, .
\end{equation}
This is an integral equation of Abel type that we can invert. As shown in \cite{PY2d}
the inversion of the equation is given by
\begin{equation}
 \psi(t)= \frac{-2}{\sqrt{1-t^2}}\left[1+\int_1^\infty \frac{\sqrt{s^2-1}}{s^2-t^2}s\psi(s) \frac{ds}{\pi} \right] \, , \qquad 0<t<1
 \label{eq:cond11} \, .
\end{equation}
Eq.~\eqref{eq:cond11} is an integral equation that determines $\psi(t)$ for
$0<t<1$ as function of $\psi(t)$ for $t>1$. Also, note that the
behavior of $\psi(t)$ near $t=1$ is
\begin{equation}
\psi(t)\sim \left(1-t^2\right)^{-1/2}\qquad \mbox{as}\qquad t\rightarrow 1^-
\end{equation}
Substituting the results of Eqs.~\eqref{eq:sq}, (\ref{eq:A})-(\ref{eq:Qq}) and
(\ref{eq:h}) into Eq.~(\ref{eq:O&Z1}) gives
\begin{eqnarray}
 \frac{1}{\tilde{Q}(-q)} &=& \left[1-\lambda \int_0^1Q(s)e^{iqs}ds\right] \\
 &\times& \left[1+\rho\left(\frac{2\pi}{q}\right)^{k+\frac{1}{2}}\int_0^\infty t^{k+\frac{3}{2}}J_{k+\frac{1}{2}}(qt)\psi(t) dt\right] \nonumber
\end{eqnarray}
Multiplying by $\exp(-iqt)$ with $t>0$ and integrating with respect
to $q$ from $-\infty$ to $\infty$ we obtain
\begin{eqnarray}
 &&2^kk! \int_0^1Q(s)\delta(s-t)ds -  \int_t^\infty (s^2-t^2)^k \psi(s) sds \\
 &+&\lambda \int_0^1ds Q(s) \int_{|t-s|}^\infty (s'^2-(s-t)^2)^k\psi(s')s'ds' = 0 \nonumber
\end{eqnarray}
which leads to
\begin{eqnarray}
 && 2^kk!Q(t)\Theta(1-t)-\int_t^\infty (s^2-t^2)^k \psi(s) sds \\
 && =-\lambda \int_0^1ds Q(s) \int_{|t-s|}^\infty (s'^2-(t-s)^2)^k\psi(s')s'ds'
 \nonumber \, ,
\end{eqnarray}
where $\Theta(x)$ is the Heaviside function. Recall that $\psi(t)$
is defined only for $t>0$. Therefore one has
\begin{widetext}
\begin{eqnarray}
2^kk!\,Q(t)-\int_t^\infty (s^2-t^2)^k \psi(s) sds &=&-\lambda \int_0^1ds \,Q(s) \int_{|t-s|}^\infty (s'^2-(t-s)^2)^k\psi(s')s'ds' \qquad 0<t<1
\label{eq:cond2}\\
\int_t^\infty (s^2-t^2)^k \psi(s) sds &=&\lambda \int_0^1ds\, Q(s) \int_{(t-s)}^\infty (s'^2-(t-s)^2)^k\psi(s')s'ds' \qquad t>1
\label{eq:cond3}
\end{eqnarray}
\end{widetext}
Replacing $\psi(t)$ for $0<t<1$ in Eq.~(\ref{eq:cond2})  with its value
as given by Eq.~(\ref{eq:cond11}) yields an equation that depends
only on $\psi(t)$ for $t>1$, namely
\begin{widetext}
\begin{equation}
 Q(t)=A(t)+\int_1^\infty B(s,t)\psi(s) sds -\lambda \int_0^1\left[A(t-s) + \int_{1}^\infty B(s',t-s)\psi(s')s'ds'\right]Q(s)ds \qquad 0<t<1
 \label{eq:cond22}
\end{equation}
\end{widetext}
with
\begin{eqnarray}
A(t)&=&-\frac{2}{(2k+1)!!}(1-t^2)^{k+\frac{1}{2}}\\
B(s,t)&=&\frac{1}{2^kk!}(s^2-t^2)^k\,I\left(\frac{s^2-1}{s^2-t^2};\frac{1}{2},\frac{1}{2}+k\right)
\end{eqnarray}
where $I(z;a,b)$ is the regularized Beta function \cite{abram}. On
the other hand, using Eq.~(\ref{eq:cond11}) and differentiating
$k$-times with respect to $t^2$ and once with respect to $t$ one can
rewrite Eq.~(\ref{eq:cond3}) as
\begin{widetext}
\begin{eqnarray}
 \psi(t) &=&-(-1)^{k}\lambda \left(\frac{d}{tdt}\right)^{k+1}\int_{t-1}^1\left[A(t-s)+\int_1^\infty B(s',t-s)\psi(s')s'ds' \right]Q(s)ds \nonumber\\
 &&-(-1)^{k}\frac{\lambda}{2^kk!} \left(\frac{d}{tdt}\right)^{k+1}\int_0^{t-1}\left[\int_{(t-s)}^\infty (s'^2-(t-s)^2)^k\psi(s')s'ds' \right] Q(s) ds\qquad 1<t<2 \label{eq:cond33}\\
 \psi(t) &=&-(-1)^{k}\frac{\lambda}{2^kk!} \left(\frac{d}{tdt}\right)^{k+1}\int_0^1\left[\int_{(t-s)}^\infty (s'^2-(t-s)^2)^k\psi(s')s'ds' \right] Q(s) ds\qquad t>2
\label{eq:cond34}
\end{eqnarray}
\end{widetext}

Our approach has reduced the PY problem for hard hyperspheres to the
solution of the set of one-dimensional integro-differential
equations (\ref{eq:cond22}),(\ref{eq:cond33}) and (\ref{eq:cond34}) for the auxiliary functions $\psi(s)$ and $Q(s)$. Once these functions have been determined, the physically relevant functions may be determined; $g(r)$ from \eqref{hdefn} and \eqref{eq:solh}, $c(r)$ from \eqref{eq:solc} and \eqref{eq:solphi}.
We note that unlike the odd dimensional case \cite{leuth1}, in even dimensions it is not possible to separate the problem of finding the direct correlation
function $c(r)$ from that of finding the pair correlation function
$g(r)$. This is because the behavior of the auxiliary function $\psi(t)$ for $0<t<1$ is
coupled to its behavior for $t>1$ through
Eq.~(\ref{eq:cond11}). Although we were unable to find an analytical
solution valid for all $\rho$, a numerical algorithm to find the
numerical solution of these equations can easily be implemented.
Before dealing with the numerical analysis, let us first consider
the equation of state in the present formulation of the problem.

\section{Equation of state}
There are two methods used to calculate the equation of state when
the radial distribution function, $g(r)$, is known. Without the
assumptions made in deriving the PY equation \cite{PY}, these two
methods would yield the same equation of state. The difference in
the equations of state calculated using these two methods therefore
provides an estimation of the error made by using the PY
approximation. The first method of calculating the equation of state
uses the isothermal compressibility $\kappa_T$ which is given
by~\cite{hansen}
\begin{equation}
 \rho \beta^{-1} \kappa_T=\frac{1}{\beta}\,\left(\frac{\partial \rho}{\partial P^{(c)}}\right)_T=s(q=0) \, ,
\end{equation}
where $\beta$ is the inverse temperature $1/k_B T$. Using
Eqs.~(\ref{eq:A}) and (\ref{eq:solcq}), it is easy to deduce that
\begin{equation}
 \frac{1}{s(0)}=1- \frac{2(2\pi)^{k}\rho}{(2k+1)!!}\int_0^1 t^{2(k+1)}\phi(t) dt
 \label{eq:s(0)} \, ,
\end{equation}
which can be further simplified by successive integration by parts to give
\begin{equation}
 \frac{1}{s(0)}=1- \frac{2\pi^k\rho}{k!} \int_0^1 dt \left[\int_t^1 (s^2-t^2)^k
 \phi(s)sds\right] \,.
\end{equation}
Using Eq.~\eqref{eq:solphi0} we may eliminate $\phi(s)$ to yield
\begin{equation}
\frac{1}{s(0)}=1-2\lambda \int_0^1 \left[Q(t)-\lambda\int_t^1 Q(s)Q(s-t)ds\right] dt
\end{equation}
Thus, the compressibility equation of state can be written as
\begin{eqnarray}
 && \frac{\beta \, P^{(c)}}{\rho} - 1=\\
 && -\frac{2}{\lambda}\int_0^{\lambda}\left\{\int_0^1 \left[Q(t)-\lambda'\int_t^1 Q(s)Q(s-t)ds\right] dt\right\}\lambda'd\lambda'\,,\nonumber
 \label{pceqn}
\end{eqnarray}
The second method to obtain the equation of state is derived from
the virial theorem and is given by
\begin{equation}
 \beta \, P^{(v)}=\rho+\frac{\pi^{k+1}}{(k+1)!}\,\frac{\rho^2}{2}\,g(1^+)
 \label{eq:Pv} \, .
\end{equation}
Finally, using Eqs.~\eqref{hdefn} and (\ref{eq:solh}),
Eq.~(\ref{eq:Pv}) becomes
\begin{equation}
 \frac{ \beta P^{(v)}}{\rho}-1=\frac{\pi\lambda}{2^{k+1}(k+1)!}\left[1+\int_1^\infty \frac{s\psi(s)}{\sqrt{s^2-1}}\frac{ds}{\pi} \right]
 \label{eq:Pv1} \, ,
\end{equation} where $\lambda$ is defined in \eqref{lambdadefn}.

\section{Numerical Procedure}

As in our earlier work \cite{PY2d}, we solve Eqs.~(\ref{eq:cond22}),(\ref{eq:cond33}) and (\ref{eq:cond34}) for the auxiliary functions $Q(s)$ and $\psi(s)$ iteratively. However, the present method of solution is simpler than
the one used previously, as explained below. We pose power series
\begin{eqnarray}
 Q(t) &=&\sum_{i=0}^\infty \lambda^i q_i(t) \, , \qquad 0<t<1 \, , \label{series1} \\
 \psi(t) &=&\sum_{i=0}^\infty \lambda^i \psi_i(t) \, , \qquad t>1 \, ,
 \label{series2}
\end{eqnarray}
for the unknown functions, and substitute these power series into
Eqs.~(\ref{eq:cond22}),(\ref{eq:cond33}) and (\ref{eq:cond34}). At
zeroth order, we obtain
\begin{equation}
 q_0(t)=A(t); \quad \psi_0(t)=0
 \label{eq:cond0n} \, .
\end{equation}
For $i \geq 0$, we find
\begin{widetext}
\begin{eqnarray}
 \psi_{i+1}(t) &=&-(-1)^{k}\left(\frac{d}{tdt}\right)^{k+1}\sum_{j=0}^i\int_{t-1}^1\left[A(t-s)\delta_{j,i}+\int_1^\infty B(s',t-s)\psi_{i-j}(s')s'ds' \right]q_{j}(s)ds \nonumber \\
 &&-\frac{(-1)^{k}}{2^kk!} \left(\frac{d}{tdt}\right)^{k+1}\sum_{j=0}^i\int_0^{t-1}\left[\int_{(t-s)}^\infty (s'^2-(t-s)^2)^k\psi_{i-j}(s')s'ds' \right]q_{j}(s)ds,\qquad 1<t<2;
 \label{eq:cond1n}\\
 \psi_{i+1}(t) &=&-\frac{(-1)^{k}}{2^kk!} \left(\frac{d}{tdt}\right)^{k+1}\sum_{j=0}^i\int_0^1\left[\int_{(t-s)}^\infty (s'^2-(t-s)^2)^k\psi_{i-j}(s')s'ds' \right]q_{j}(s)ds,\qquad t>2 ;
 \label{eq:cond2n}\\
 q_{i+1}(t)&=&\int_1^\infty B(s,t)\psi_{i+1}(s) sds -\sum_{j=0}^i\int_0^1\left[A(t-s) \delta_{j,i}+ \int_{1}^\infty B(s',t-s)\psi_{i-j}(s')s'ds'\right]q_{j}(s)ds,\; 0<t<1
 \label{eq:cond3n} .
\end{eqnarray}
\end{widetext}

We use an iterative procedure, starting with \eqref{eq:cond0n}, to
calculate successively $\psi_i(t)$ and then $q_i(t)$ using
\eqref{eq:cond1n}-\eqref{eq:cond3n}. In the above formulation of the
problem, the only difficulty is that one has to differentiate
$(k+1)$-times. However, this is balanced by the fact that the
integrals to be computed have no singular behavior, in contrast with
our earlier method of solution for the case $k=0$ (i.e.~$d=2$)\cite{PY2d}.

Rather than computing the equation of state for various densities
$\rho$ (as in earlier work\cite{lado}), we compute the virial
coefficients, $B_i$, which are defined by
\begin{equation}
\beta \, P=\sum_{i=1}^\infty B_i\rho^i
 \label{eq:virial} \, .
\end{equation}
Thanks to the iterative procedure presented above, these
coefficients, namely $B_i^{(c)}$, from the
compressibility route \eqref{pceqn}, and $B_i^{(v)}$, from the virial route \eqref{eq:Pv1}, are directly given by the numerical resolution of the problem.

The first two coefficients are identical and may be computed analytically:
\begin{equation}
 B_1^{(c)}=B_1^{(v)}=1
 \label{eq:B1} \, ,
\end{equation}
and
\begin{equation}
 B_2^{(c)}=B_2^{(v)}=\frac{\pi^{k+1}}{2(k+1)!}
 \label{eq:B2} \, .
\end{equation} Higher order coefficients must be found numerically. For $i\geq1$ we find
\begin{eqnarray}
 B_{i+2}^{(c)} &=& -\frac{2(2\pi)^{k(i+1)}}{i+2}
 \label{virialc} \\
 &\times& \int_0^1 \left[q_{i}(t)-\sum_{j=0}^{i-1}\int_t^1 q_j(s)q_{i-1-j}(s-t)ds\right] dt \nonumber
 \, .
\end{eqnarray}
and
\begin{equation}
 B_{i+2}^{(v)} = \frac{(2\pi)^{k(i+1)}}{2^{k+1}(k+1)!}\int_1^\infty \frac{s\psi_{i}(s)}{\sqrt{s^2-1}}ds
 \label{virialv} \, ,
\end{equation}

The calculation of the first two virial coefficients via the two
routes (c and v) produce identical results  (coincident with the
exact results) because the PY approximation is exact up to first
order \cite{rohrmann08} in $\rho$. What is not so evident is that
for the same reason both the third virial coefficient and the first
order pair correlation function, $g_1(r)$, are also reproduced
exactly by the PY theory. $B_3$ is given in closed form by\cite{B3}
\begin{equation}
 B_3^{(c)}=B_3^{(v)}=B_3^{exact} = \frac{\pi^{2k+2}}{2[(k+1)!]^2} \frac{B_{3/4}(k+\frac{3}{2},\frac{1}{2})}{B(k+\frac{3}{2},\frac{1}{2})}
 \label{eq:B3} \, ,
\end{equation}
where $B(a,b)=\Gamma(a)\Gamma(b)/\Gamma(a+b)$ is the beta function,
$B_x(a,b)$ is the incomplete beta function \cite{abram}. Note that
for each integer dimension this expression can be written in a simpler form
without the need for transcendental functions. However, it is not
possible to write a simpler general form. Furthermore, $g_1(r)$ is given by
\begin{equation}
 g_1(r) = 1+\int_r^\infty \frac{t\psi_1(t)}{\sqrt{t^2-r^2}} \frac{dt}{\pi} = \Theta(r-1) \left[1+2^d \alpha_2(r;1) \right]
 \label{eq:psi1} \, ,
\end{equation}
where  $\alpha_2(r;1)$ is the scaled intersection volume \cite{Torquato02,Torquato06}.

It is also worthwhile mentioning that $B_4$ was recently\cite{B4}
evaluated exactly for all dimensions up to $d=12$, though a
closed-form formula for any $d$ is not available. We reproduce the
values of $B_4$ for even dimensions $d\leq8$, of interest here, in
Table~\ref{B4}.

\begin{table}
\centering \setlength{\tabcolsep}{0.8 em}
\begin{tabular}{ccc}
$d$ & $B_4^{exact}$ & $B_4^{Numercial}$\\
\hline
2 &$\frac{\pi^3}{8}  \left(2-\frac{9 \sqrt{3}}{2 \pi} +\frac{10}{\pi ^2}\right)$&$2.062$\\
4 &$\frac{\pi^6}{64} \left(2-\frac{27 \sqrt{3}}{4 \pi}+\frac{832}{45 \pi ^2}\right) $&$2.281$\\
6 &$\frac{\pi^9}{1728} \left(2-\frac{81 \sqrt{3}}{10 \pi }+\frac{38848}{1575 \pi ^2}\right)$&$0.576$\\
8  &$\frac{\pi^{12}}{110592} \left( 2-\frac{2511 \sqrt{3}}{280 \pi }+\frac{17605024}{606375 \pi ^2}\right)$& $-0.021$\\
 \end{tabular}
\caption{The exact expressions for $B_4$ as well as their
corresponding numerical values for some even dimensions obtained
recently \cite{B4}.}
 \label{B4}
\end{table}


We note that the computation of the correlation function $g(r)$ and
the equation of state do not present any significant difficulties.
The integrands that must be computed in the calculation of the
coefficients $B_i^{(v)}$, see  \eqref{virialv}, and $g(r)$, see
\eqref{eq:solh}, have weak square root singularities that may be
dealt with by integration by parts or by subtraction of the
singularity.  We validate our numerical implementation of the
iterative procedure outlined above by comparing our results with the
earlier results for the case $k=0$ (i.e.~$d=2$)\cite{PY2d}.

\section{Numerical Results}

In this section, we present the results of our numerical
computations in dimensions $d=4,6,8$. A brief description of our
numerical scheme is presented in Appendix \ref{app:num}.

\subsection{Comparison with Monte-Carlo simulations}

We begin by comparing the pair correlation function, $g(r)$,
obtained from our solution of the PY equation with that obtained in
the Monte-Carlo (MC) simulations of others\cite{whitlock}. For
simplicity we present here numerical results obtained using the
first ten terms of the series \eqref{series1}-\eqref{series2} and a
spatial resolution $\dx=10^{-3}$. These are representative of the
results obtained by taking more terms and using a higher spatial
resolution.

\begin{figure}
\centering
\includegraphics[height=7cm]{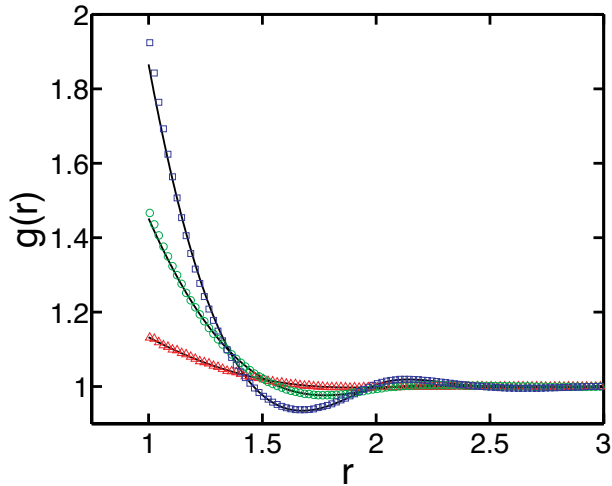}
\caption{(Color online) The correlation function $g(r)$ computed
from the PY equation (curves) and from Monte Carlo
simulations\cite{whitlock} (symbols) in $d=4$. Results are plotted
for $\rho=0.1$ ($\triangle$), $\rho=0.3$ ($\bigcirc$) and $\rho=0.5$
($\square$). The PY results are plotted taking $10$ terms of the
series \eqref{series1}-\eqref{series2} and were computed with
$\dx=10^{-3}$.} \label{d=4mc}
\end{figure}

\begin{figure}
\centering
\includegraphics[height=7cm]{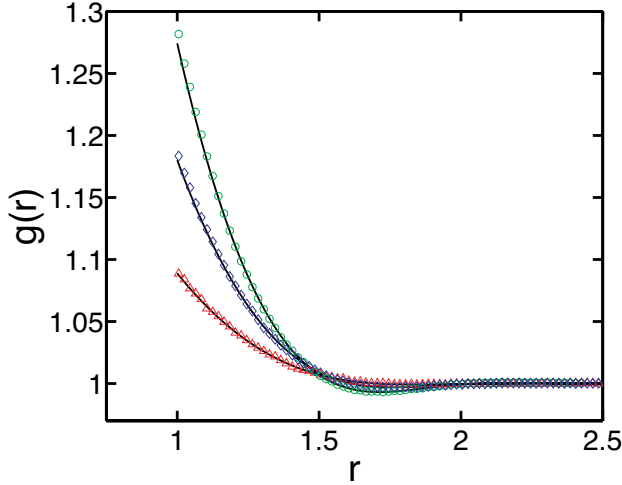}
\caption{(Color online) The correlation function $g(r)$ computed
from the PY equation (curves) and from Monte Carlo
simulations\cite{whitlock} (symbols) in $d=6$. Results are plotted
for $\rho=0.1$ ($\triangle$), $\rho=0.2$ ($\Diamond$) and $\rho=0.3$
($\bigcirc$). The PY results are plotted taking $10$ terms of the
series \eqref{series1}-\eqref{series2} and were computed with
$\dx=10^{-3}$.} \label{d=6mc}
\end{figure}

\begin{figure}
\centering
\includegraphics[height=7cm]{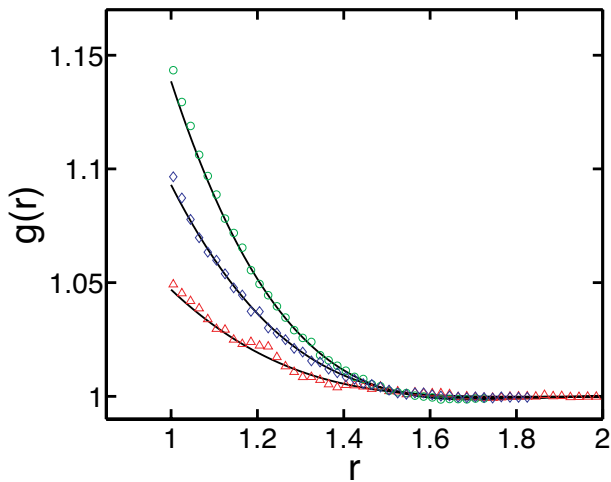}
\caption{(Color online) The correlation function $g(r)$ computed
from the PY equation (curves) and from Monte Carlo
simulations\cite{whitlock} (symbols) in $d=8$. Results are plotted
for $\rho=0.1$ ($\triangle$), $\rho=0.2$ ($\Diamond$) and $\rho=0.3$
($\bigcirc$). The PY results are plotted taking $10$ terms of the
series \eqref{series1}-\eqref{series2} and were computed with
$\dx=10^{-3}$.} \label{d=8mc}
\end{figure}

Figures \ref{d=4mc}, \ref{d=6mc} and \ref{d=8mc} show $g(r)$ for
$d=4,6$ and $8$, respectively with several different values of
$\rho$. (The values of $\rho$ used here are chosen to be smaller
than the radius of convergence of the virial series, see \S
\ref{virial} and \S \ref{conv} below. For larger values of $\rho$,
the series \eqref{series1}-\eqref{series2} does not converge, though
it may perhaps be resummed to improve the convergence for larger
densities.) We see that there is generally very good agreement in
each case between the PY results (solid curves) and MC simulations
(symbols). We note that, as expected, the discrepancy is largest for
$r\approx1$ --- similar to the behavior observed in other
dimensions\cite{lado,PY2d}. Interestingly, this discrepancy seems to
diminish as the dimension increases. The good agreement between the
MC results and our solutions of the PY equation shows that our
method of solution works well.

\subsection{Virial coefficients\label{virial}}

We computed the virial coefficients using the two expressions
\eqref{virialc} and \eqref{virialv} for $d=2$ as a check of our
numerical scheme. (Note that this is not a trivial verification since the
numerical scheme is completely different to that used
previously\cite{PY2d}.) We obtain the same results as reported
previously \cite{PY2d}, though our estimation of errors leads to
some slightly different values for the last digit of the higher
order coefficients.

The virial coefficients calculated by this method for $d=4,6$ and
$8$ are reproduced in tables \ref{coeffs4d}-\ref{coeffs8d}. These
tables show the virial coefficients resulting from both the virial
$(v)$ and compressibility $(c)$ routes, as well as the virial
coefficients obtained by earlier MC calculations\cite{clisby06}.
Note that all the virial coefficients up to $B_4$ are known exactly
(see Eq.~\eqref{eq:B3} and Table~\ref{B4} above), and agree with the
MC calculations. In general, we see that $B_i^{(c)}$ is a better
estimator of the true virial coefficient, $B_i^{(MC)}$, than
$B_i^{(v)}$.

\begin{table}
\centering \setlength{\tabcolsep}{0.8 em}
\begin{tabular}{cccc}
$i$ & $B_i^{(v)}$ & $B_i^{(c)}$& $B_i^{(MC)}$\\
\hline
3 &$3.083$&$3.083$&$3.083$\\
4 &$1.774$&$2.466$& $2.281$\\
5 &$0.988$&$1.602$&$1.323$\\
6  &$0.315$& $0.875$&$0.707$\\
7  &$0.285$& $0.463$&$0.323$\\
8 & $-0.103$ & $0.193$&$0.161$\\
9   &$0.278$& $0.120$& $0.061$\\
10 &$-0.386$&$0.0046$&$0.038$\\
11 &$0.642$&$0.071$& ---\\
12 &$-1.058$&$-0.076$& ---\\
13 &$1.793$&$0.134$& ---\\
14 &$-3.088$&$-0.210$& ---\\
15 &$5.402$&$0.348$& ---
 \end{tabular}
\caption{Numerical values of the first fifteen virial coefficients
in four dimensions. The  $B_i^{(MC)}$ are the results from Monte
Carlo calculations presented previously~\cite{clisby06}. $B_i^{(v)}$
and $B_i^{(c)}$ are the values found from the solution to the PY
equation using Eqs.~\eqref{virialv} and \eqref{virialc}
respectively.} \label{coeffs4d}
\end{table}

\begin{table}
\centering \setlength{\tabcolsep}{0.8 em}
\begin{tabular}{cccc}
$i$ & $B_i^{(v)}$ & $B_i^{(c)}$& $B_i^{(MC)}$\\
\hline
3 &$2.276 $&$2.276 $&$2.276 $\\
4 &$0.189$&$0.795$& $0.576 $\\
5 &$0.51(7)$&$0.342$&$0.335 $\\
6  &$-0.72(2)$& $-0.104$&$-0.200$\\
7  &$1.43(7)$& $0.278$&$0.389 $\\
8 & $-3.0(4)$ & $-0.509$&$-0.688$\\
9   &$6.8(6)$& $1.043$& $1.326 $\\
10 &$-16.(3)$&$-2.255$&$-2.696$\\
11 &$40.(5)$&$5.097$& ---\\
12 &$-10(2)$&$-11.94(7)$& ---\\
13 &$26(5)$&$28.86(1)$& ---\\
14 &$-7(10)$&$-71.51(5)$& ---\\
15 &$19(15)$&$181.09(1)$& ---
 \end{tabular}
\caption{Numerical values of the first fifteen virial coefficients
in six dimensions. The  $B_i^{(MC)}$ are the results from Monte
Carlo calculations presented previously~\cite{clisby06}. $B_i^{(v)}$
and $B_i^{(c)}$ are the values found from the solution to the PY
equation using Eqs.~\eqref{virialv} and \eqref{virialc}
respectively.} \label{coeffs6d}
\end{table}

\begin{table}
\centering \setlength{\tabcolsep}{0.8 em}
\begin{tabular}{cccc}
$i$ & $B_i^{(v)}$ & $B_i^{(c)}$& $B_i^{(MC)}$\\
\hline
3 &$0.966$&$0.966 $&$0.966 $\\
4 &$-0.13(1)$&$0.065$& $-0.021 $\\
5 &$0.21(3)$&$0.074$&$0.126 $\\
6  &$-0.30(2)$& $-0.078$&$-0.155 $\\
7  &$0.5(1)$& $0.11(5)$&$0.239 $\\
8 & $-0.9(0)$ & $-0.19(0)$&$-0.406 $\\
9   &$1.(7)$& $0.33(6)$& $0.747 $\\
10 &$-3.(4)$&$-0.6(4)$&$-1.466 $\\
11 &$7.(5)$&$1.(3)$& ---\\
 \end{tabular}
\caption{Numerical values of the first eleven virial coefficients in
eight dimensions. The  $B_i^{(MC)}$ are the results from Monte Carlo
calculations presented previously~\cite{clisby06}. $B_i^{(v)}$ and
$B_i^{(c)}$ are the values found from the solution to the PY
equation using Eqs.~\eqref{virialv} and \eqref{virialc}
respectively.} \label{coeffs8d}
\end{table}

We seem to see the same trends as Rohrmann \emph{et
al.}\cite{rohrmann08} regarding the way in which the PY virial
coefficients bound the true virial coefficient for $d\geq7$ (though
our results reveal that the transition in behavior they observe
occurs already at $d = 6$, rather than $d=7$). In particular, we
see that $B_i^{(v)}<B_i<B_i^{(c)}$ for even $i>4$ and
$B_i^{(v)}>B_i>B_i^{(c)}$ for odd $i>5$.

Note also the intermediate behavior observed
previously\cite{rohrmann08} for $d=5$ is observed already with
$d=4$, though the details are a little different: we find that
$B_i<B_i^{(c)}<B_i^{(v)}$ for odd $i>7$ and
$B_i>B_i^{(c)}>B_i^{(v)}$ for even $i>8$ --- though one would need
to see more real virial coefficients to say this with more
confidence.

\subsection{Convergence of virial series\label{conv}}

\begin{figure}
\centering
\includegraphics[height=7cm]{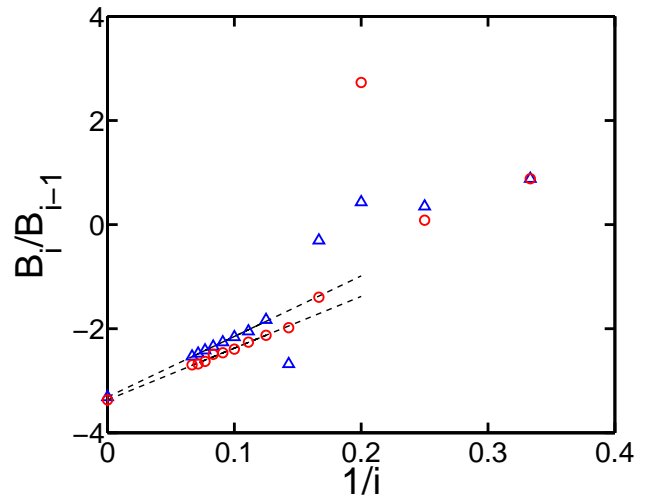}
\caption{(Color online) Domb-Sykes plot for the virial coefficients
in six dimensions. The ratios are based on the virial coefficients
for $P^{(c)}$ ($\triangle$) and $P^{(v)}$ ($\bigcirc$). The dashed
lines show the relevant linear fits at large $i$ and the points at
the intersection with the $y$-axis shows the extrapolated limits and
the associated error bars.} \label{ds_6d}
\end{figure}

A question of considerable interest is the radius of convergence of
the virial series. This question is closely related to the nature of
the singularity closest to the origin, which was addressed recently
in odd dimensions \cite{rohrmann08}. An estimate of this radius of
convergence may be made by using the Domb-Sykes
plot\cite{hinch,vandyke}. This is an extension of the ratio test in
which the ratio between successive terms, $B_i/B_{i-1}$ is plotted
as a function of $1/i$. Such a plot often reveals a linear
behavior, the intercept of which then provides an estimate of the
radius of convergence of the series, $\rho_{\mathrm{conv}}$, via
\begin{equation}
\rho_{\mathrm{conv}}^{-1}=\lim_{i\rightarrow\infty}\left|\frac{B_i}{B_{i-1}}\right|
\, .
\end{equation}
An example of such a plot is shown in Fig.~\ref{ds_6d} for the
virial coefficients in six dimensions.

\begin{figure}
\centering
\includegraphics[height=7cm]{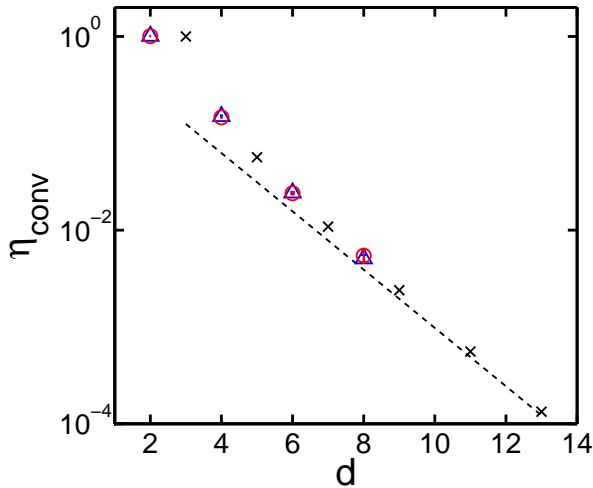}
\caption{(Color online) The radius of convergence for the virial
series, $\eta_{\mathrm{conv}}$, as a function of dimension, $d$. The
radius of convergence in odd dimensions ($\times$) is taken from the
work of Rohrmann \emph{et al.}\cite{rohrmann08}. The results in even
dimensions are found from the relevant Domb-Sykes plots and show
estimates based on $P^{(c)}$ ($\triangle$) and $P^{(v)}$
($\bigcirc$). The dashed line shows the relationship
$\eta_{\mathrm{conv}}=2^{-d}$, which is believed to describe the
large $d$ behavior of $\eta_{\mathrm{conv}}$.} \label{rad_conv}
\end{figure}

We used the Domb-Sykes plot to determine the radius of convergence
of the series\footnote{For $d=4$ it was necessary to take the first
twenty virial coefficients to find a sufficiently linear trend to
warrant extracting the intercept.} for $d=2,4,6$ and $8$. The
results are given in table \ref{radii} as the radius of convergence
for the series in $\eta$, $\eta_{\mathrm{conv}}$, along with
estimates of the error in each case (Eq.~\eqref{eq:rho-eta} may be
used to convert $\eta_{\mathrm{conv}}$ to $\rho_{\mathrm{conv}}$).
These results are also plotted in Fig.~\ref{rad_conv} and combined
with the results for odd dimensions $d\leq13$ obtained by the
analysis of Rohrmann \emph{et al.}\cite{rohrmann08}. These results
seem to confirm the assertion that as $d\rightarrow\infty$,
\begin{equation}
\eta_{\mathrm{conv}}\sim 2^{-d}
 \label{largedasy} \, ,
\end{equation}
which was conjectured by Frisch and Percus \cite{frisch} for
the full problem. The Domb-Sykes plots for $d=4,6,8$ have negative
intercepts with the vertical axis and positive slopes there (see,
Fig.~\ref{ds_6d}, for example). This shows that, in these cases, the
singularity that limits the radius of convergence of the virial
series is a branch point on the negative real
axis\cite{vandyke,hinch}. In contrast, the Domb-Sykes plot for two
dimensions suggests that the relevant singularity is a pole on the
positive real axis.

\begin{table}
\centering \setlength{\tabcolsep}{0.8 em}
\begin{tabular}{ccc}
$d$ & $\eta^{(v)}_{\mathrm{conv}}$ & $\eta^{(c)}_{\mathrm{conv}}$\\
\hline
2 & $1.008\pm0.002$ & $1.01\pm0.002$\\
4 & $0.146\pm 8\times10^{-5}$ & $0.150\pm 0.003$\\
6 & $0.024\pm 7\times10^{-4}$ & $0.024\pm 6\times10^{-4}$ \\
8 & $0.0055\pm8\times10^{-4}$ & $0.0051\pm5\times10^{-4}$
 \end{tabular}
\caption{The radii of convergence for the virial series, for both
the virial (v) and the compressibility (c) routes, as a function of
the dimension $d$.} \label{radii}
\end{table}

\section{Discussion}
In this paper we generalized a semi-analytic method to solve the PY
equation for hard discs \cite{PY2d} to the case of hard hyperspheres
in even dimensions. The essence of this approach is a reduction of
the PY equation to a set of integro-differential equations
for two auxiliary functions $Q(s)$ and $\psi(s)$ as given by
Eqs.~(\ref{eq:cond22}),(\ref{eq:cond33}) and (\ref{eq:cond34}). The
correlation functions and the equation of state can be determined
easily from these auxiliary functions. We suggest an efficient
iterative numerical method to solve these equations and determine
the auxiliary functions in $d=4,6$ and $8$. Using this method we are
able to determine the values of the virial coefficients within the
PY approximation and compare them with the first ten virial
coefficients for the full problem \cite{clisby06}. We also obtain
results for the pair correlation function which compare well with
the available MC simulations\cite{whitlock}.

The principal advantage of this approach is that it provides
directly the virial series, and so it yields the equation of state
for all values of $\rho$ at the same time provided that
$\rho<\rho_{\mathrm{conv}}$. This is in contrast with other
approaches where each value of $\rho$ requires a separate
calculation\cite{lado}. An important consequence is that we can
study the convergence of the series. We have shown that the virial
series predicted by the PY theory for hyperspheres in even
dimensions has a branch point on the negative real axis
$\eta=-\eta_{\mathrm{conv}}$ for all $d\geq4$. This is observed in
both the virial and compressibility routes to determining the
equation of state, similar to what happens in PY in odd dimensions
\cite{rohrmann08}. The position of this singularity as a function of
the the dimension $d$ is also consistent with the conjecture by
Frisch and Percus \cite{frisch} for the full problem in large
dimensions. The successful prediction of the singularity supports
the idea that the PY theory  approaches the exact problem as the
dimensionality increases. An important conclusion from this
discussion is that semi-phenomenological equations of state, such as
the generalizations of the celebrated Carnahan-Starling equation of
state \cite{CarStar} (for three-dimensional hard spheres) to higher
dimensions \cite{Song}, are inferior to the PY theory since they are
not able to reproduce such a branch point singularity. This characteristic is also missing from far more elaborated equations of state \cite{Luban90,robles}.

Another interesting prediction of the current work is that the exact
virial coefficients $B_i$ for $d=4$ may change sign if they are
computed for sufficiently large $i$. We observe that in PY theory with $d=4$ the virial coefficients determined by both routes are negative for even $i\geq12$. This suggests that the exact coefficient may also become negative for sufficiently large $i$ and will hopefully motivate the calculation of further coefficients using the methods of Clisby and McCoy\cite{clisby06}. More
generally the question of negative values of the virial coefficients
in various dimensions is pertinent\cite{clisby04b,clisby05,rohrmann08}.

As explained above, this method allows, in principle, calculations
to arbitrary precision, and it could be interesting to obtain more
virial coefficients by doing so. Such progress might also allow for a
comparison of the pair correlation function with Molecular Dynamics
simulations \cite{Lue05,Lue06,Skoge06} at large densities.

It could also be interesting to apply the approach developed here to
polydisperse mixtures \cite{Lebowitz,Baxter70,Gonzalez}, to sticky
hard spheres (i.e., hard spheres with an adhesive short range
interaction) \cite{Baxter68} and to much higher dimensions.

\subsection*{Acknowledgments}

We would like to thank Profs Whitlock and Bishop for sharing
their data with us, and Prof. Santos for his useful comments.
This work was supported by the Royal Commission for the Exhibition
of 1851 (D.V.). Laboratoire de Physique Statistique is associated
with Universities Paris VI and Paris VII.

\begin{appendix}
\section{Numerical scheme\label{app:num}}

We solve the system of equations \eqref{eq:cond0n}-\eqref{eq:cond3n}
by discretizing in space using steps of size $\dx$. Integrals are
calculated using the trapezoidal rule, which is first order
accurate. To calculate derivatives we developed two different
schemes: one which used forward differencing (first-order accuracy)
and the other using central differencing (second-order accuracy).
The results obtained with these two differentiation schemes are
consistent with one another. Iterations proceed from $i=0$ using the
values for $\psi_0(t)$ and $q_0(t)$ from \eqref{eq:cond0n} to
determine $\psi_1(t)$.  The function $\psi_1(t)$ may then be used
with the discretized version of \eqref{eq:cond3n} to determine
$q_1(t)$. This process is then repeated $N$ times, corresponding to
determining the first $N$ terms in the series expansions of
$\psi(t)$ and $q(t)$. The results presented in this paper typically
have $N=10$.

\begin{figure}
\centering
\includegraphics[height=7cm]{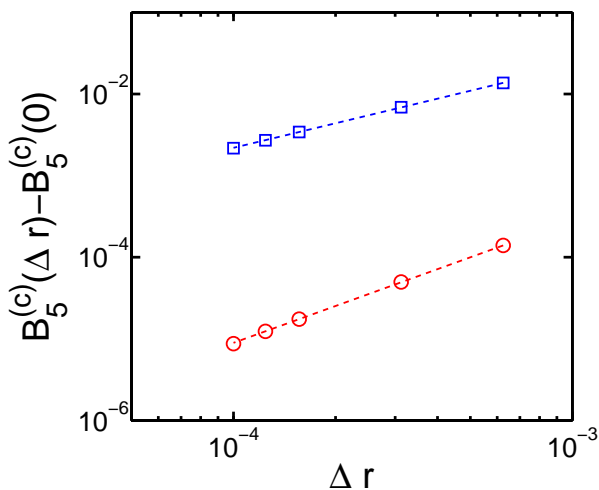}
\caption{(Color online) An example of the convergence of the virial
coefficients as a function of $\dx$ for $B_5^{(c)}$ with $d=6$ using
the two numerical schemes described in the text: forward
differencing ($\square$) and central differencing ($\bigcirc$). For
both data sets the dashed lines are the result of the fitting
procedure, with a slope of $1$ for forward differencing, and with a
slope of $3/2$ for central differencing. Note that the plot is on a
log-log scale.}
 \label{dr_conv}
\end{figure}

The virial coefficients were determined by numerical integration,
again using the trapezoidal rule. To determine the values of these
coefficients more accurately, we performed computations at different
spatial resolutions, i.e.~ with different values of $\dx$. Plotting
the behavior of these numerically determined coefficients as a
function of $\dx$ we then extrapolated the observed trend to
$\dx=0$. The observed trend was linear for the scheme using
differentiation by forward differencing, as expected since both
discretizations involve errors of order $\dx$. The second scheme
(using differentiation by central differencing) is more complicated
since the first order error of integration is mixed with a
second-order error in differentiation. Typically with this scheme we
find errors that scale like $\dx^{3/2}$. In Fig.~\ref{dr_conv} below
we show an example of the convergence using these two schemes.

To determine the values presented in tables
\ref{coeffs4d}-\ref{coeffs8d} we use the values extrapolated to
$\dx=0$ from both numerical schemes. An estimate of the error
introduced in this extrapolation procedure was obtained using the
$95\%$ confidence interval for the value at $\dx=0$. The results
presented in tables \ref{coeffs4d}-\ref{coeffs8d} are correct to $3$
decimal places or as otherwise indicated.

\end{appendix}

\end{document}